\documentclass[review,3p]{elsarticle}%

\usepackage[T1]{fontenc}
\usepackage[utf8]{inputenc} 								
\usepackage[english]{babel} 								
\usepackage{lmodern}

\usepackage{amsthm}
\usepackage{amssymb}
\usepackage{amsmath}
\usepackage{mathtools}
\usepackage{textcomp}
\usepackage[overload]{empheq}

\usepackage{epsfig}
\usepackage{subfig}
\usepackage[dvipsnames]{xcolor}
\usepackage[colorlinks]{hyperref}

 \usepackage{lineno}

\journal{Advances in Water Resources}

\usepackage{bm}
\newcommand{\W}{\bm{W}}
\newcommand{\St}{\bm{S}}
\newcommand{\A}{\bm{\mathcal{A}}}
\newcommand{\Ap}{\bm{A}}
\newcommand{\Ie}{\bm{I}_{m}}
\newcommand{\Fr}{\mathtt{Fr}}
\newcommand{\nc}{\mathtt{nc}}
\newcommand{\Eg}{\mathtt{E}_\varphi}
\newcommand{\LL}{\mathbf{\Lambda}}				
\newcommand{\RR}{\mathtt{\mathbf{R}}}			
\newcommand{\D}{\bm{\mathcal{D}}}

\newenvironment{sistema} 
{\left\lbrace\begin{array}{@{}l@{}}}%
{\end{array}\right.}

\usepackage{booktabs}       								
\usepackage{siunitx}												
\usepackage{multirow}												

\usepackage[]{changes}
\definechangesauthor[name={Francesco}, color=Green]{F}
\definechangesauthor[name={Alessandro}, color=red]{A}
\definecolor{bblue}{rgb}{0.23,0.4,0.7}
\definechangesauthor[name={Valerio}, color=bblue]{val}

\setremarkmarkup{[#2]}

\begin{document}

\begin{frontmatter}



\title{Efficient analytical implementation of the DOT Riemann solver for the \\de Saint Venant-Exner morphodynamic model}


\author[unife]{F. Carraro}
\author[unife]{A. Valiani}
\author[unife]{V. Caleffi}

\address[unife]{Department of Engineering, University of Ferrara, Italy}

\begin{abstract}
Within the framework of the de Saint Venant equations coupled with the Exner equation for morphodynamic evolution, this work presents a new efficient implementation of the Dumbser-Osher-Toro (DOT) scheme for non-conservative problems.
The DOT path-conservative scheme is a robust upwind method based on a complete Riemann solver, but it has the drawback of requiring expensive numerical computations.
Indeed, to compute the non-linear time evolution in each time step, the DOT scheme requires numerical computation of the flux matrix eigenstructure (the totality of eigenvalues and eigenvectors) several times at each cell edge.
In this work, an analytical and compact formulation of the eigenstructure for the de Saint Venant-Exner (dSVE) model is introduced and tested in terms of numerical efficiency and stability.
Using the original DOT and PRICE-C (a very efficient FORCE-type method) as reference methods, we present a convergence analysis (error against CPU time) to study the performance of the DOT method with our new analytical implementation of eigenstructure calculations (A-DOT).
In particular, the numerical performance of the three methods is tested in three test cases: a movable bed Riemann problem with analytical solution; a problem with smooth analytical solution; a test in which the water flow is characterised by subcritical and supercritical regions.
For a given target error, the A-DOT method is always the most efficient choice.
Finally, two experimental data sets and different transport formulae are considered to test the A-DOT model in more practical case studies.
\end{abstract} 


\end{frontmatter}


\section{Introduction}
\label{sec:intro}
Morphodynamic models are used to predict river and nearshore evolution describing the interactions between sediments and water flow.
Due to the importance of many engineering applications requiring morphodynamic predictions, great attention has been devoted to the development of more efficient numerical techniques.
Indeed, the constant increase of numerical scheme efficiency and computer power allows the application of the mathematical models to even long-term and complex phenomena.

Many mathematical models used in the study of natural river, estuarine and nearshore morphodynamics \cite{Berthon2012,Caleffi2007,Canestrelli2009,Carraro2016,Kelly2010,Postacchini2014} are based on the de Saint Venant equations (also called shallow water equations, SWE) coupled with the Exner sediment continuity equation \cite{Exner1925}.
This set of balance laws has the main disadvantage of containing a non-conservative product due to the variable bed topography in the momentum balance \cite{Pares2006a}.

From the physical point of view, when the Froude number is small, morphodynamics combines two processes characterized by very different time scales.
The bed evolution is much slower than the hydrodynamics and the interaction between the two processes is weak \cite{MASSPEED}.
Therefore, most morphodynamic models for practical applications were developed by splitting the computation of the hydrodynamics from the bed topography adaptation (e.g.\ \cite{Brunner2016,Lesser2004,Martini2004,Vetsch2017}).
These models separately solve the SWE for a fixed bed topography, later updating such topography by using the Exner equation with the updated hydrodynamic quantities.
In this manner the hydrodynamic part of the governing system can be treated as a strictly hyperbolic system of equations, thus avoiding the complexities typical of non-conservative balance laws.

In spite of the wide application of this decoupled approach, the importance of integrating the whole system of PDE in a coupled manner has been highlighted in the recent years \cite{Caleffi2007,Cordier2011}, both to better represent the physical phenomenon in the case of intense sediment transport and to avoid numerical instabilities.
It is important to note that rapidly-varying hydraulic conditions, like sudden hydraulic transients and steady or unsteady breaking waves, are the typical conditions requiring a coupled approach.
For example, studying the morphodynamics of the nearshore zone, \citet{Kelly2010} show that the use of a decoupled approach in presence of bore-driven sediment transport leads to the overestimation of the net offshore transport in the swash zone.
\citet{Postacchini2014} show that the erosion of the bed is significantly larger in the uncoupled models if a swash forced by a dam-break is considered.

Moreover, \citet{Cordier2011} showed that decoupling hydrodynamics and morphodynamics could generate some non-physical oscillations related to the hyperbolicity of the system of equations.
These oscillations occur even when a robust and well-balanced numerical scheme for the de Saint Venant system is used, especially when the bed evolution is very dynamic \cite{Cordier2011}.
In some cases, such oscillations can be avoided by reducing the Courant–Friedrichs–Lewy coefficient \cite{Toro2009a}, forcing a smaller time step in time-integration and decreasing the numerical efficiency.

However, the fully coupled approach also presents a critical issue. 
The non-conservative form of the de Saint Venant-Exner (dSVE) system prevents formulation of the classic Rankine-Hugoniot relationship, leading to uncertainties about the mathematical representation of the hydraulic jumps. See \cite{Abgrall2010} for an analysis of this issue.

A theoretical contribution to the solution of this problem comes from the Dal Maso, LeFloch, and Murat (DLM) theory \cite{dal1995definition} which introduces the generalized Rankine-Hugoniot relationship for a given arbitrary integration path.
The DLM theory does not give information about the selection of the path, leaving some degrees of freedom.
This freedom was used by \citet{Pares2006a} to introduce the family of path-conservative (or path-consistent) methods to properly handle non-conservative system of equations.

Formally consistent numerical solutions can be obtained by coupling the path-conservative schemes with appropriated paths. Furthermore, by selecting the path on the basis of physical considerations, a target specific behaviour of the resulting numerical models can be achieved.
For example, to develop a well-balanced scheme with respect to the case of still water with a constant free surface elevation, simple segment paths are sufficient \cite{Castro2007}. Otherwise, to obtain energy-balanced scheme that exactly reproduce the constancy of the total head in the case of steady flows \cite{Murillo2013}, more elaborated paths must be introduced \cite{Caleffi2017,Caleffi2016,Fjordholm2011,Valiani2017}.

In the last ten years, different methods have been based on the path-conservative scheme \cite{Canestrelli2009,Castro2006,Dumbser2011}.
In particular, a very general and robust method is the Dumbser-Osher-Toro (DOT) Riemann solver \cite{Dumbser2011}. 
It is an upwind scheme based on a complete Osher-type Riemann solver that performs well with various non-conservative problems (e.g.\ \cite{Caleffi2016,Dumbser2009,Muller2014}).
Nevertheless, DOT has the drawback of requiring several numerical computations of the full eigenstructure of the flux matrix per time step.
Thus the method has remained unattractive to parts of the computational community due to its high computational cost, which makes the method more suited for basic numerical tests than for practical engineering applications.

The path-conservative approach can also be applied in a centred (or symmetric) numerical framework.
In this context, a recent and efficient scheme is the PRICE-C of \citet{Canestrelli2009}, a coupled scheme of the FORCE type.
Therefore, among the always monotone methods, it assigns the smallest numerical viscosity within the Riemann fan \cite{Toro2009a,Toro2001}.
An upwinding of the fully coupled PRICE-C method is implemented in GIAMT2D \cite{Siviglia2013}, an open source numerical model developed for research aims but also suitable for practical engineering problems.

In \cite{shekariI2013}, the DOT and the PRICE-C schemes are used to solve the Single Pressure Model (SPM) for the isothermal compressible two-phase flows. 
This non-conservative mathematical model has not simple analytical expressions for its eigenvalues.
The comparison between the two schemes is performed in terms of numerical efficiency and accuracy of the corresponding solutions, in particular for the challenging simulation of near single phase flows.
In general, for the same level of accuracy, the DOT scheme shows a better numerical efficiency with respect to the PRICE-C scheme.
This is due to the excessive spurious dissipation of the PRICE-C central scheme.
Conversely, the DOT scheme gives less accurate results in the simulation of near single phase flows compared to the PRICE-C scheme.

The DOT and the PRICE-C schemes are also compared in \cite{Stecca2016} where both the approaches are applied to the de Saint-Venant-Hirano model for mixed-sediment morphodynamics, which is another non-conservative mathematical model without analytical expressions for eigenvalues and eigenvectors. 
Also in this framework the PRICE-C centred method shows a larger numerical dissipation with respect to the DOT upwind method.
For this application, the DOT method results computationally inefficient. In fact, because the sediment mixture is discretized into several fractions, the de Saint-Venant-Hirano model becomes a large system of equations with an eigenstructure very cumbersome to be computed numerically.

In a preliminary work about the numerical efficiency of the DOT schemes \cite{Carraro2016} the eigenstructure is computed numerically, analytically and using an approximate solution based on a perturbative analysis.
To test these different approaches a suitable set of test cases is considered in \cite{Carraro2016} without comparing the results with those of different numerical schemes or with experimental data.
Furthermore, in \cite{Carraro2016} a systematic and strictly quantitative convergence analysis (error vs CPU time) is missing.

Taking the original DOT and PRICE-C methods as reference, in this work we present an analytical implementation of the DOT solver for the one-dimensional de Saint Venant-Exner model.
A compact and easy to implement analytical formulation of the dSVE eigenstructure is used to improve the numerical efficiency and to simplify the model coding.
This avoids complex numerical tools to compute eigenvalues, eigenvectors and inverse matrices, which must be calculated several times at each time step.
For conciseness, hereafter the DOT method with the numerical computation of the eigenstructure replaced by such new analytical solution is denoted A-DOT.

The specific purpose of this optimization is to make more convenient the DOT solver for practical morphodynamic problems; hence it can be considered a step towards the implementation of a fully coupled morphodynamic model for a 1D and 2D river and coastal simulations. 
From the practical point of view, the A-DOT method can give a significant gain of computational time in long-term morphodynamic simulations.
Moreover, this tool does not require the tuning of case-dependent parameters, so that its application becomes straightforward, once analytical preliminaries are established.

This paper is organised as follows:
in section \ref{Model} the one-dimensional de Saint Venant-Exner mathematical model is summarized;
in section \ref{sec:pathCon} the structure of the path-conservative scheme is summarized, with a focus on the formulation of three different jump functions (also called fluctuations);
in section \ref{sec:results} a computational cost analysis is presented for three suitable test cases;
in section \ref{sec:experimental} the A-DOT numerical model is tested against two different experimental data sets;
in section \ref{sec:conclusion} conclusions are drawn.

\section{Mathematical Model}
\label{Model}
In the present work, the morphodynamic problem of a wide rectangular cross-section is described by the classical unit-width de Saint Venant equations coupled with the Exner equation. When the total sediment transport is supposed to match the sediment transport capacity of the flow, the resulting dSVE model can be written as
\begin{subequations}
	\begin{align}[left = \empheqlbrace\,]
		&\frac{\partial h}{\partial t} + \frac{\partial q}{\partial x} = 0 \label{eq:SV-Ex1}\\
		&\frac{\partial q}{\partial t} + \frac{\partial}{\partial x}\left(\frac{1}{2}g h^2 + \frac{q^2}{h}\right) + g h\,\frac{\partial z}{\partial x}=   -g h\, s_f \label{eq:SV-Ex2}\\
		&\frac{\partial z}{\partial t} + \xi\frac{\partial q_s}{\partial x} = 0 \label{eq:SV-Ex3}
	\end{align}
	\label{eq:SV-Ex}
\end{subequations}

\noindent with
$h(x,t)$ the water depth;
$q(x,t)$ the specific water discharge;
$z(x,t)$ the bed elevation;
$g$ the gravitational acceleration;
$s_f$ the friction slope;
$q_s$ the sediment transport specific discharge (in volume);
$\xi=1/(1-p)$ where $p$ is the porosity of the riverbed.
In this work $p$ is assumed constant, and thus also $\xi$ is constant.

It is well known that the de Saint Venant-Exner model contains a non-conservative product, that is the bed topography term in the momentum equation \eqref{eq:SV-Ex2}.
Therefore, to properly include the non-conservative product \cite{dal1995definition}, system \eqref{eq:SV-Ex} must be written in a quasi-linear form as
\begin{equation}
	\frac{\partial \W}{\partial t}+\A(\W)\frac{\partial \W}{\partial x} = \St(\W)\,,
	\label{eq:quasiLinNC}
\end{equation}
in which $\W$ is the vector of the conservative variables, $\A(\W)$ is the flux matrix and $\bm{S}(\W)$ is the vector of the source terms
\begin{equation}
	\W= \begin{bmatrix} h \\	q \\ z \end{bmatrix},  \quad
	\A(\W)=
	\begin{bmatrix}
											0 								&										1										&			0	  		\\
									 c^2-u^2							&	 								 2u										&		 c^2			\\
		\xi\frac{\partial q_s}{\partial h} 	&	 \xi\frac{\partial q_s}{\partial q}		&			0
	\end{bmatrix}, \quad
	\St(\W)= \begin{bmatrix} 0 \\	-c^2 s_f \\ 0 \end{bmatrix}\,,
	\label{eq:W&A_definition}
\end{equation}
with
$u=q/h$ the depth averaged velocity and
$c=\sqrt{gh}$ the propagation celerity of gravitational waves.

System \eqref{eq:SV-Ex} is composed by three partial differential equations in five unknowns: $h(x,t)$, $q(x,t)$, $z(x,t)$, $s_f(x,t)$ and $q_s(x,t)$.
Therefore, other two relations are required to close the system.
The friction term is provided by a classical closure, namely
\begin{equation}
s_f=\frac{q^2}{{K_s}^2\,h^{10/3}}\,,
\end{equation} 
where $K_s$ is the Strickler coefficient.

For the sake of simplicity, the Grass closure formula \cite{Grass1981} could be chosen for the computation of bed load discharge:
\begin{equation}
	q_s = A_g \, u^3\,,
\label{eq:Grass}
\end{equation}
where $A_g$ is a constant that depends on the grain properties of the bed.
Once $q_s$ is expressed by Eq.~\eqref{eq:Grass}, system \eqref{eq:SV-Ex} is always strictly hyperbolic \cite{Cordier2011}.

The proposed DOT Riemann solver is not actually related to a specific closure for the evaluation of $q_s$.
As soon as the hyperbolicity of system \eqref{eq:SV-Ex} is verified (see \cite{Cordier2011} for the conditions that must be respected), the results described in this paper remain valid for any expression of the sediment discharge, which allows the evaluation of $q_s$ derivatives.
In particular, in section \ref{subs:Toro} three different formulation of $q_s$ are considered to compare numerical solution and experimental data:
\begin{enumerate}
	\item a simple power-law similar to the Grass formula, but with a threshold on the incipient transport velocity:
	\begin{equation}
		q_s = A_g \, (u-u_{cr})^3\,;
	\label{eq:GrassThr}
	\end{equation}
	\item the Meyer-Peter \& M\"uller (MPM) formula \cite{Meyer-Peter1948},
	\begin{equation}
		q_s = \Phi(\theta) \sqrt{g\,\left(S_g-1\right)\,{d_{50}}^3},\quad \text{with}\quad
		\Phi(\theta) = \begin{cases} 8(\theta-0.047)^{3/2} & \mbox{if } \theta>0.047 \\ 0 & \mbox{otherwise} \end{cases}
	\label{eq:MPM}
	\end{equation}
	where $S_g = 2.6$ is the relative density of the sediment grain, $d_{50}$ is the median sediment size and $\theta$ is the Shields stress given by
	\begin{equation}
	\theta = \frac{s_f\,h}{\left(S_g-1\right)\,{d_{50}}}\,;
	\label{eq:Shields}
	\end{equation}
	\item the Van Rijn formula \cite{VanRijn1984},
	\begin{equation}
		q_s = 0.053\frac{\mathtt{T\,}^{2.1}}{{\mathtt{D}_*}^{0.3}} \sqrt{g\,\left(S_g-1\right)\,{d_{50}}^3}
		,\quad \text{with}\quad \mathtt{T}=\frac{{u_*}^2-{u_*}^2_{cr}}{{u_*}^2_{cr}}\quad\text{and}\quad\mathtt{D_*}=d_{50}\left[\frac{g(S_g-1)}{\nu}\right]^{1/3},
	\label{eq:vanRijn}
	\end{equation}
	where $\nu = 10^{-6}$ m$^2$/s is the cinematic viscosity, $u_*=\sqrt{g\,h\,s_f}$ is the shear stress velocity and ${u_*}_{cr}$ is the critical shear stress velocity given by:
	\begin{equation}
		{u_*}_{cr} = \theta_{cr}\sqrt{g(S_g-1)d_{50}} \qquad \text{with}\qquad \theta_{cr} = 0.03\;.
	\label{eq:u_star}
	\end{equation}
\end{enumerate}

\section{Path-conservative Finite Volume Schemes}
\label{sec:pathCon}
The presence of the non-conservative term in equation \eqref{eq:SV-Ex2} prevents computation of a flux function for implementation of a classic finite volume scheme.
Implementation of a path-conservative (or path-consistent) scheme is an established technique to handle such non-conservative term \cite{Castro2006,Pares2006a}.

A first order of accuracy path-conservative scheme reads as
\begin{equation}
\W^{n+1}_i = \W_i^n - \frac{\Delta t}{\Delta x}\left(\D_{i+\frac{1}{2}}^- + \D_{i-\frac{1}{2}}^+\right) + \Delta t\,\St\left(\W^n_i\right) \,,
\label{eq:scheme}
\end{equation}
in which: $\W^{n+1}_i$ and $\W^n_i$ are the cell-averaged values of the vector $\W$ at time levels $t^{n+1}$ and $t^n$, respectively;
$\Delta t\,\St(\W_i^n)$ is the integral of the source term within the $i$-$th$ cell, between time $t^n$ and $t^{n+1}$;
$\D^\pm_{i\pm\frac{1}{2}}$ are the jump functions (also called fluctuations), with the compatibility condition on the cell edges
\begin{equation}
\D_{i+\frac{1}{2}}^- + \D_{i+\frac{1}{2}}^+ = \int_0^1{\A\left(\Psi(\W_{i+\frac{1}{2}}^-,\W_{i+\frac{1}{2}}^+,s)\right)\frac{\partial\Psi}{\partial s}\mathrm{d}s}
\label{eq:jumps}
\end{equation}
with $\Psi(\W_{i+\frac{1}{2}}^-,\W_{i+\frac{1}{2}}^+,s)$ the path connecting the left state $\W_{i+\frac{1}{2}}^-$ and the right state $\W_{i+\frac{1}{2}}^+$ across the \mbox{$(i+\frac{1}{2})$-$th$} edge between two adjacent cells, with $s\in[0,1]$.

For the sake of simplicity, from now on the only considered path is the segment path, given by
\begin{equation}
	\Psi(s)=\W_{i+\frac{1}{2}}^-+s(\W_{i+\frac{1}{2}}^+-\W_{i+\frac{1}{2}}^-).
\label{eq:path}
\end{equation}

Working at the first order of accuracy allows a very simple treatment of the source term.
Thus, in this section we focus only on the computation of the fluctuation terms $\D$.
However, more general approaches should have been considered for higher order of accuracy or stiff hyperbolic balance laws \cite{Dumbser2009,Dumbser2008a,Siviglia2013}.

\subsection{The Original DOT Riemann Solver}
\label{sub:DOT}
A robust and general method to compute the jump function $\D$ is an Osher type Riemann solver: the DOT (Dumbser-Osher-Toro) Riemann solver \cite{Dumbser2011}.

According to this method, the general jump function can be expressed as
\begin{equation}
		\D^\pm_{i+\frac{1}{2}} = \frac{1}{2}  \int_0^1{\left[\A\left(\Psi \left(\W^-_{i+\frac{1}{2}}, \W^+_{i+\frac{1}{2}}, s \right)\right)\pm \left|\A\left(\Psi \left(\W^-_{i+\frac{1}{2}}, \W^+_{i+\frac{1}{2}}, s \right)\right) \right|\right] \frac{\partial\Psi}{\partial s} \mathrm{d}s}\,.
	\label{eq:DOTjump}
\end{equation}
Using a G-point quadrature rule and the segment path \eqref{eq:path}, the discrete formulation of Eq.~\eqref{eq:DOTjump} becomes
\begin{equation}
	\D^\pm_{i+\frac{1}{2}} = \frac{1}{2} \,\sum_{k=1}^G{ w_{Gk}\left(\A\left(\Psi (s_k)\right) \pm \left|\A\left(\Psi (s_k)\right)\right|\right)}\,\left(\W^+_{i+\frac{1}{2}} - \W^-_{i+\frac{1}{2}}\right) \;,
	\label{eq:DOTjumpN}
\end{equation}
where $w_{Gk}$ and $s_k$ are weights and nodes of the Gauss quadrature and the usual convention for the absolute value of a matrix is applied:
\begin{equation}
	\left| \A\right| = \RR \left| \LL\right| \RR^{-1}, \quad \left|\LL\right|=\text{\textbf{diag}}(|\lambda_1|,\cdots,|\lambda_n|)\,,
	\label{eq:asbA}
\end{equation}
where $\RR$ is the matrix of right eigenvectors of $\A$,
$|\LL|$ is the diagonal matrix of the eigenvalues in absolute value,
and $\RR^{-1}$ is the inverse of $\RR$ (i.e.\ the matrix of the left eigenvectors).

The DOT solver preserves several important properties of the Osher Riemann solver.
The DOT jump is an entropy-satisfying, non-linear function.
Moreover, as shown by \citet{Dumbser2011}, using the full eigenstructure of the hyperbolic system, DOT attributes an individual (and generally different) numerical viscosity to each characteristic field, and particularly to the intermediate fields.
In this sense, it is a complete Riemann solver.
This feature allows accurate simulation of phenomena strongly influenced by the intermediate waves, such as morphodynamic evolution.

In fact, the de Saint Venant-Exner model \eqref{eq:SV-Ex} has three characteristic variables: the characteristic variable related to the smallest eigenvalue (in absolute value) can be mainly associated with the bed elevation and the others with the water flow \cite{MASSPEED,Lyn2002}.
Moreover, the intermediate Riemann wave in the dSVE model is related to the same smallest eigenvalue.
Thus, a sharp treatment of the intermediate wave is mandatory for precise reproduction of the bed elevation.

Nevertheless, as can be deduced by Eq.~\eqref{eq:DOTjumpN}, the DOT solver requires computation of the $\A$ matrix eigensystem a large number of times for each time step ($G \times (\nc+1)$, where $\nc$ is the number of cells) to perform the integration along the path: the numerical computation of this eigensystem makes the DOT solver an expensive procedure from the point of view of computational cost.

\subsection{The PRICE-C method}
\label{sub:PRICE}
PRICE-C \cite{Canestrelli2012} is an easy to implement FORCE-type method  \cite{Toro2009a} optimized for the de Saint Venant-Exner model.
As an evolution of the original PRICE-T \cite{Toro2003}, it adopts one of the most efficient formulation of the jump function for centred schemes.
Formally, it writes
\begin{equation}
		\D^\pm_{i+\frac{1}{2}} = \Ap^\pm_{i+\frac{1}{2}}\,\left(\W^+_{i+\frac{1}{2}} - \W^-_{i+\frac{1}{2}} \right)\,,
	\label{eq:PRICEjump}
\end{equation}
with
\begin{equation}
		\Ap^\pm_{i+\frac{1}{2}} = \frac{1}{4}\,\left[2\,\Ap_{i+\frac{1}{2}}^{\bm{\Psi}} \pm \frac{\Delta x}{\Delta t}\Ie \pm \frac{\Delta t}{\Delta x}\left( \Ap_{i+\frac{1}{2}}^{\bm{\Psi}}\right)^2 \right]\,.
\label{eq:AiPRICE}
\end{equation}
In \eqref{eq:AiPRICE}, $\Ap_{i+\frac{1}{2}}^{\bm{\Psi}}$ is given by a simple integration along the path:
\begin{equation}
		\Ap_{i+\frac{1}{2}}^{\bm{\Psi}} = \int_0^1{\A\left(\Psi \left(\W^-_{i+\frac{1}{2}}, \W^+_{i+\frac{1}{2}}, s \right)\right)\, \mathrm{d}s}\,;
\label{eq:Apsi}
\end{equation}
while $\Ie$ is the modified identity matrix
\begin{equation}
	\Ie = \begin{bmatrix}
					1 & 0 & 0 \\
					0 & 1 & 0 \\
					0 & 0 & \epsilon 
				\end{bmatrix}\,,
\label{eq:Im}
\end{equation}
where $\epsilon$ is a small parameter that prevents undesirable diffusion of the bed forms \cite{Canestrelli2009}.
In general, the $\epsilon$ value depends on the specific problem and can be estimated as the ratio between the slowest and the fastest celerities of the Riemann fans \cite{Siviglia2013}.
In this work, for each test the $\epsilon$ value has been chosen with a trial and error procedure.

Unlike an upwind approach, a centred scheme does not explicitly use wave propagation information contained in the hyperbolic system of equations \cite{Toro2009a}.
This feature allows the use of a jump function that does not require computation of the flux matrix eigenstructure in the development of the PRICE-C method, resulting in easy implementation and computational efficiency.
The drawback is the great diffusivity of the method in terms of bed elevation, requiring introduction of the $\epsilon$ parameter to control the numerical viscosity associated with the bed evolution, for which a general case-independent value is not available.

\subsection{Definition of the new A-DOT formulation}
\label{sub:ADOT}
The main drawback of the DOT solver is the numerical computation of $|\A|$ as defined in \eqref{eq:asbA}.
As with Eqs.\ \eqref{eq:DOTjump} and \eqref{eq:DOTjumpN}, its computation is required $G \times (\nc+1)$ times for each time step along the entire time evolution.
Thus, with a focus only on the dSVE model, an analytical formulation of $|\A|$ is introduced here.

The characteristic polynomial of the system matrix, defined in \eqref{eq:W&A_definition}, is obtained as usual, with the imposition that $|\A(\W)-\lambda \bm{I}|=0$.
When this polynomial is divided by $c^3$ to obtain a non-dimensional formulation, it becomes
\begin{equation}
	\frac{\lambda^3}{c^3}-2\,\Fr\,\frac{\lambda^2}{c^2}-\left(1 - \Fr^2 + \xi\frac{\partial q_s}{\partial q} \right)\,\frac{\lambda}{c} - \xi\frac{\partial q_s}{\partial h}\frac{1}{c}=0.
	\label{eq:charPol}
\end{equation}
According to the method proposed by \citet{cardano1545}, in the form applied in \cite{Valiani2008}, the three solutions of Eq.~\eqref{eq:charPol}, i.e.\ the three non-dimensional eigenvalues, can be analytically expressed as:
\begin{subequations}
	\label{eq:lambda}
	\begin{align}
		\frac{\lambda_1}{c}&=\frac{2}{3}\,\mathtt{Fr}-\frac{2}{3}\sqrt{k_1}\,\cos{\left(\frac{\phi}{3}-\frac{\pi}{3}\right)}\,,		 \\
		\frac{\lambda_2}{c}&=\frac{2}{3}\,\mathtt{Fr}-\frac{2}{3}\sqrt{k_1}\,\cos{\left(\frac{\phi}{3}+\frac{\pi}{3}\right)}\,,		 \\
		\frac{\lambda_3}{c}&=\frac{2}{3}\,\mathtt{Fr}+\frac{2}{3}\sqrt{k_1}\,\cos{\left(\frac{\phi}{3}\right)}\,;									
	\end{align}
\end{subequations}
where the parameters $\phi$ and $k_1$ are:
\begin{align}\label{eq:parameters}
	\phi &=\arccos{\left(\frac{k_2}{\sqrt{4{k_1}^3}}\right)}\;; \\
	k_1 &= 3+\Fr^2+3\,\xi\frac{\partial q_s}{\partial q}\;;
\end{align}
with $k_2$ given by:
\begin{equation}\label{eq:parameters_bis}
k_2 = -2\Fr^3 +18\mathtt{Fr}\left(1+\xi\frac{\partial q_s}{\partial q}\right) + 27\,\xi\frac{\partial q_s}{\partial h}\frac{1}{c}\;.
\end{equation}

\begin{figure}%
	\centering
	\includegraphics[width=0.5\columnwidth]{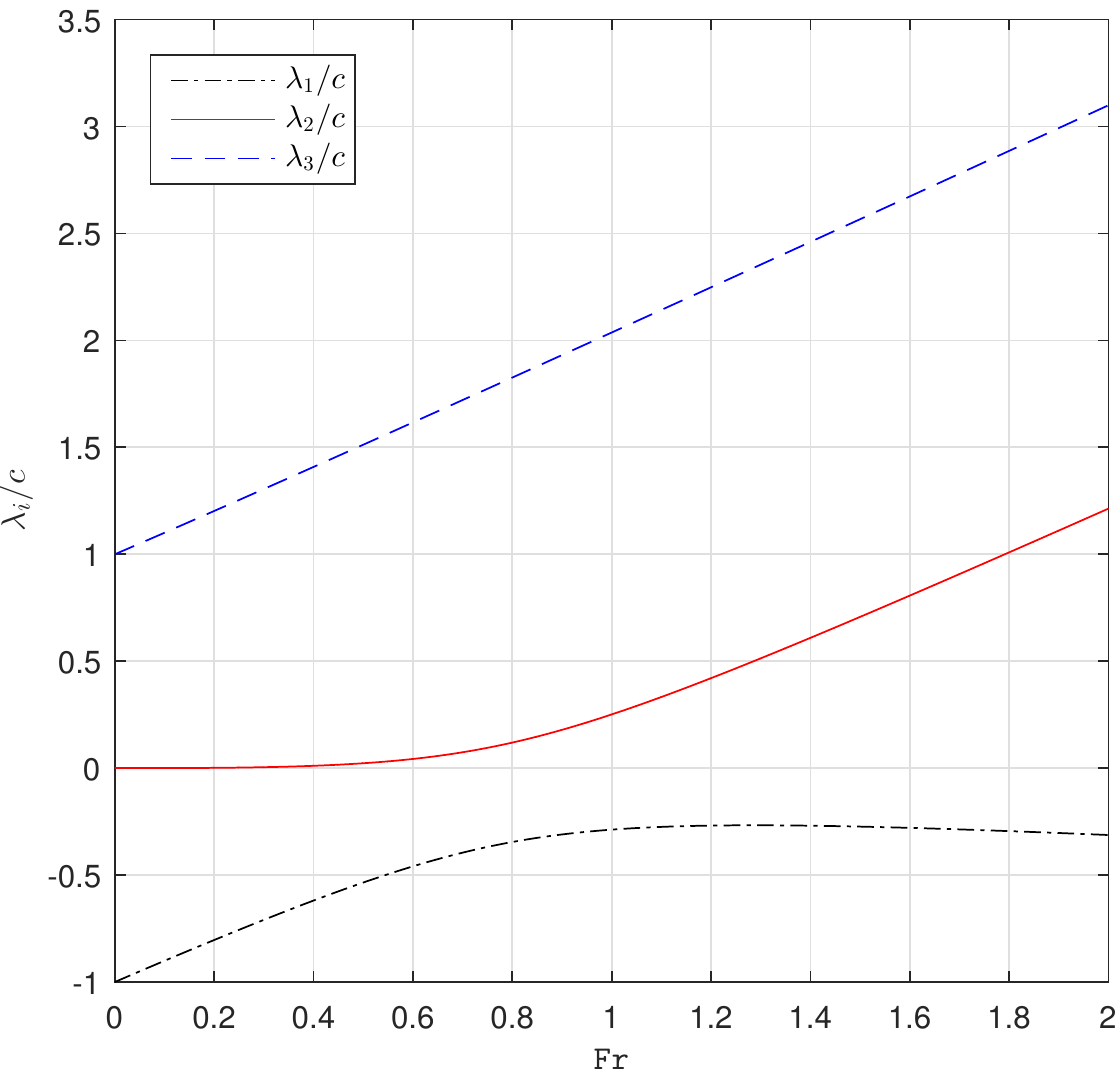}%
	\caption{Non-dimensional analytical eigenvalues, Eqs.~\eqref{eq:lambda}, with the derivatives of $q_s$ computed according to Eq.~\eqref{eq:Grass} with $A_g=0.005$ s$^2$/m.}%
	\label{fig:eig}%
\end{figure}

In Figure \ref{fig:eig}, the analytical eigenvalues \eqref{eq:lambda} are shown for $0<\Fr<2$, the derivatives of $q_s$ being computed according to Eq.~\eqref{eq:Grass} with $A_g=0.005$ s$^2$/m.
It is worth noting that the eigenvalues in Eqs.~\eqref{eq:lambda} are always real when the sediment transport is described by a power law formula, as Eq.\ \eqref{eq:Grass} or \eqref{eq:GrassThr}.
However, when the domain of definition of the parameters in \eqref{eq:parameters} is studied for a general formulation of $q_s$, the conditions of strict hyperbolicity for the dSVE system are re-obtained as defined in \cite{Cordier2011}.

Once the eigenvalues are determined, the associated right eigenvectors $\mathtt{\mathbf{r}}_i$ (as columns) can be obtained directly from their definition (up to a constant), i.e., $\A\,\mathtt{\mathbf{r}}_i = \lambda_i\,\mathtt{\mathbf{r}}_i$.
This linear system gives the matrix $\RR$ of the right eigenvectors, which can be expressed as a function of the eigenvalues as
\begin{equation}
	\RR=\left[ \mathtt{\mathbf{r}}_1, \mathtt{\mathbf{r}}_2, \mathtt{\mathbf{r}}_3 \right]
		,\quad 
		\mathtt{\mathbf{r}}_i = \left[1,\; \lambda_i,\;\frac{\left(u-\lambda_i\right)^2}{c^2}-1 	\right]^\mathtt{T}.
	\label{eq:RR}
\end{equation}
Similarly, with some simple algebraic manipulations (omitted here for brevity), the inverse of matrix $\RR$ can be written in compact form as
\begin{equation}
	\RR^{-1}=\left[ \mathbf{l}_1, \mathbf{l}_2, \mathbf{l}_3 \right]^\mathtt{T}
		,\quad
		\mathtt{\mathbf{l}}_i = \left[\frac{c^2-u^2+\lambda_j\,\lambda_k}{\left(\lambda_i-\lambda_j\right)\left(\lambda_i-\lambda_k\right)},\; \frac{2u-\lambda_{j}-\lambda_{k}}{\left(\lambda_i-\lambda_j\right)\left(\lambda_i-\lambda_k\right)},\; \frac{c^2}{\left(\lambda_i-\lambda_j\right)\left(\lambda_i-\lambda_k\right)}	\right]
	\label{eq:RR-1}
\end{equation}
where $\mathtt{\mathbf{l}}_i$ are the left eigenvectors, while $i$, $j$ and $k$ are the indexes of a circular permutation of 1, 2, and 3: 
\begin{equation}
j = \begin{cases} i+1 & \mbox{if } i<3 \\ i-2 & \mbox{otherwise} \end{cases}\qquad
k = \begin{cases} i+2 & \mbox{if } i<2 \\ i-1 & \mbox{otherwise} \end{cases}\qquad\text{with }\;i = 1,2,3\,.
\label{eq:circPerm}
\end{equation}
In Eq.~\eqref{eq:RR-1} the three row vectors $\mathtt{\mathbf{l}}_i$ are composed column-wise to obtain the square matrix $\RR^{-1}$.

According to Fig.~\ref{fig:eig}, for the dSVE system with movable bed the three eigenvalues are always distinct and real, and thus the corresponding eigenvectors are also always distinct and real.
These properties of the eigenvalues assure the well-posedness of the relationship \eqref{eq:RR-1}.
In particular, the denominators of Eq.~\eqref{eq:RR-1} are always different from zero.

By substitution in \eqref{eq:asbA} of \eqref{eq:lambda}, \eqref{eq:RR} and \eqref{eq:RR-1}, the DOT jump functions for the dSVE model can be computed algebraically.
All the elements involved in computation of $|\A|$ are defined and the proposed procedure can be summarised by the following five steps:
\begin{enumerate}
	\item compute the derivatives of the sediment discharge (e.g.\ by using Eq.\ \eqref{eq:Grass} to compute $q_s$)
	\[	q_s = A_g\left(\frac{q}{h}\right)^3\quad\Rightarrow\quad\frac{\partial q_s}{\partial h}=-3\,\frac{q_s}{h}\quad\text{and}\quad
	\frac{\partial q_s}{\partial q}=3\,\frac{q_s}{q}\,;	\]
	\item compute parameters $k_1,\,k_2,\,\phi$ according to Eqs.~\eqref{eq:parameters} and Eq.~\eqref{eq:parameters_bis};
	\item compute the eigenvalues according to Eqs.~\eqref{eq:lambda} and store them in the numerical variables $\lambda_1,\,\lambda_2,\,\lambda_3$;
	\item compute right and left eigenvectors according to Eqs.~\eqref{eq:RR} and \eqref{eq:RR-1} by using the stored eigenvalues (both to avoid typing errors and for formulation clarity);
	\item compute $|\A|$ by the definition \eqref{eq:asbA}, as the product of three matrices
	\begin{equation}
	\begin{split}
	|\A| = 
	&\begin{bmatrix}
				1			&			1			&			1			\\
		\lambda_1 &	\lambda_2 & \lambda_3 \\
		\frac{\left(u-\lambda_1\right)^2}{c^2}-1 & \frac{\left(u-\lambda_2\right)^2}{c^2}-1 & \frac{\left(u-\lambda_3\right)^2}{c^2}-1
	\end{bmatrix}
	\begin{bmatrix}
			|\lambda_1|	&	0			&	0	\\
					0 &		|\lambda_2| &  	0\\
					0&		0		&		|\lambda_3|
	\end{bmatrix}
	\\
	&\quad\quad\quad\quad\quad\quad\quad\quad\quad\quad\quad\quad\quad\quad\quad
	\begin{bmatrix}
	\frac{c^2-u^2+\lambda_2\,\lambda_3}{\left(\lambda_1-\lambda_2\right)\left(\lambda_1-\lambda_3\right)} &
	\frac{2u-\lambda_{2}-\lambda_{3}}{\left(\lambda_1-\lambda_2\right)\left(\lambda_1-\lambda_3\right)} &
	\frac{c^2}{\left(\lambda_1-\lambda_2\right)\left(\lambda_1-\lambda_3\right)}\\
	\frac{c^2-u^2+\lambda_3\,\lambda_1}{\left(\lambda_2-\lambda_3\right)\left(\lambda_2-\lambda_1\right)} &
	\frac{2u-\lambda_{3}-\lambda_{1}}{\left(\lambda_2-\lambda_3\right)\left(\lambda_2-\lambda_1\right)} &
	\frac{c^2}{\left(\lambda_2-\lambda_3\right)\left(\lambda_2-\lambda_1\right)}\\
	\frac{c^2-u^2+\lambda_1\,\lambda_2}{\left(\lambda_3-\lambda_1\right)\left(\lambda_3-\lambda_2\right)} &
	\frac{2u-\lambda_{1}-\lambda_{2}}{\left(\lambda_3-\lambda_1\right)\left(\lambda_3-\lambda_2\right)} &
	\frac{c^2}{\left(\lambda_3-\lambda_1\right)\left(\lambda_3-\lambda_2\right)}\\
	\end{bmatrix}
	\end{split}
	\label{eq:ExAbsA}\,.
	\end{equation}
\end{enumerate}

\section{Efficiency study}
\label{sec:results}
In this section, the efficiency of the proposed approaches is studied in terms of CPU time versus normalised root square errors.
For each test, at the end of the simulation, we compute the normalised root square errors for the conservative variables $h$, $q$ and $z$ as
\begin{equation}
	\Eg = \frac{\sqrt{\sum\limits_{k=1}^{\nc}{\big(\varphi(x_k,t_{\mathtt{end}}) -\varphi^\mathtt{ref}(x_k,t_{\mathtt{end}})\big)^2}}}{\sqrt{\sum\limits_{k=1}^{\nc}{{\varphi^\mathtt{ref}}(x_k,t_{\mathtt{end}})^2}}}\quad\text{with:}\quad \varphi = h,\,q,\,z.
\label{eq:Eg}
\end{equation}
The $\Eg$ values are plotted against the CPU times measured by performing several simulations and increasing the number of cells, $\nc$, within the computational grids.
Then a comparison of the solutions provided by the three approaches (PRICE-C, DOT, A-DOT) is shown.
The spatial resolutions are chosen to obtain similar CPU times for all the simulations.

To better perform the efficiency comparison, at this stage we neglect the bed friction in the momentum balance.
From a physical point of view, this assumption introduces an inconsistency because the sediment transport is strictly related to the bed shear stress and therefore to the bed friction.
Conversely, neglecting the bed friction, a tighter validation of the model is possible by considering exact analytical reference solutions.

\subsection{Test 1: a movable-bed Riemann problem}
\label{sub:murillo}
\begin{figure}[tp]%
	\centering
	\includegraphics[width=1\columnwidth]{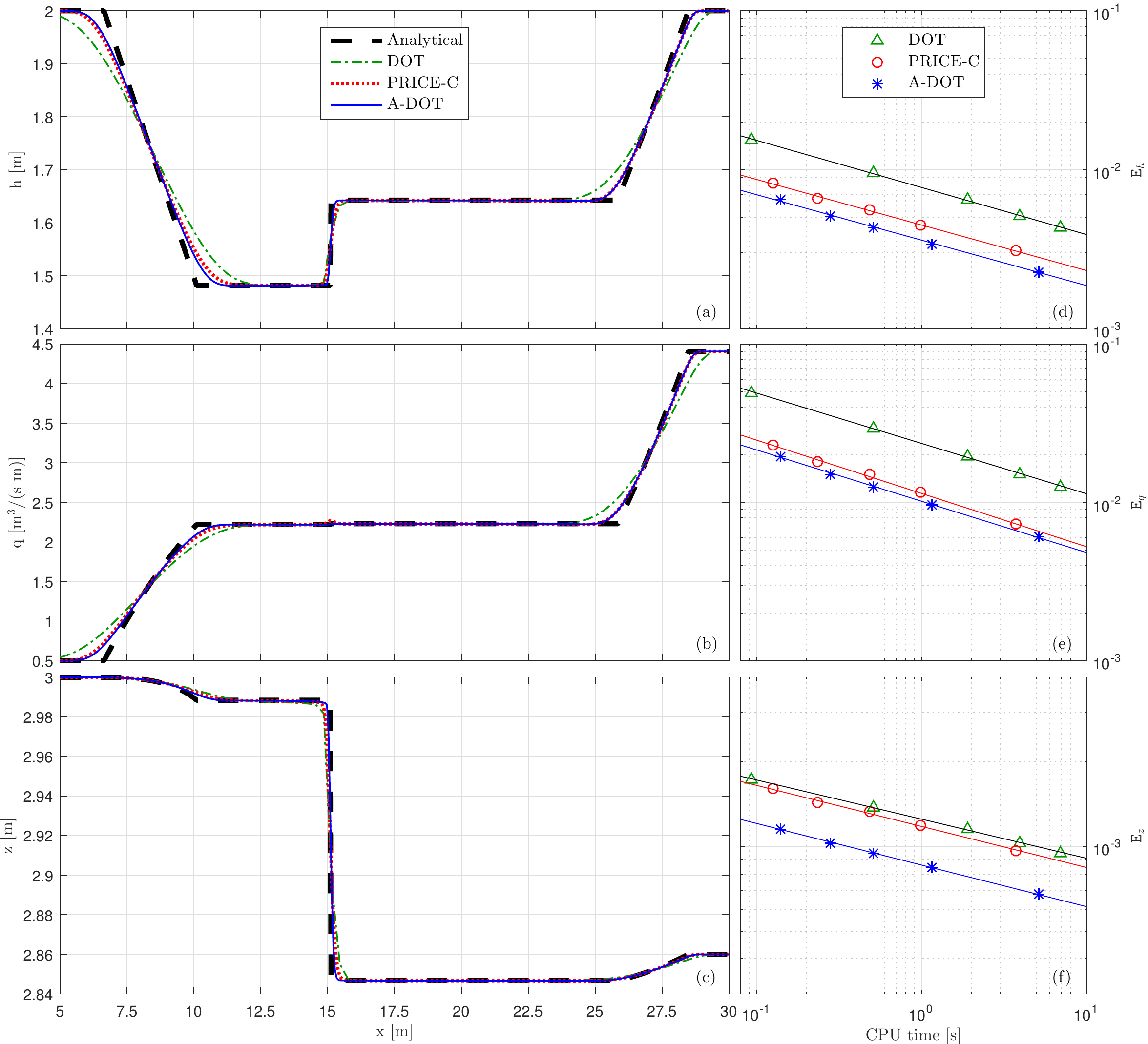}%
	\caption{Results of Test 1. Panels \emph{a, b} and \emph{c}: comparison between analytical and numerical solutions (for $h$, $q$ and $z$, respectively) at the end of the simulation (with grid size $\mathtt{nc}=$ 100 for the DOT scheme; $\mathtt{nc}=$ 500 for the PRICE-C and A-DOT schemes).  Only the solutions between $x>$ 5 m and $x<$ 30 [m] are shown. Panels \emph{d, e} and \emph{f}: CPU time versus $\Eg$, Eq.~\eqref{eq:Eg}, with $\varphi = h,\,q$ and $z$, respectively, by using: $\nc=[100, 250, 500, 750, 1000]$ for DOT; $\nc=[500, 750, 1000, 1500, 3000]$ for PRICE-C and A-DOT.}%
	\label{fig:murillo}%
\end{figure}
The first test is the comparison of the numerical solution of a movable-bed Riemann problem with the corresponding analytical solution \cite{Murillo2010}, commonly used as benchmark \cite[e.g.][]{Carraro2016}.
Given a 30 m-long straight channel, the initial conditions are imposed as two constant states: one for the left portion ($x\leq15$ m) and one for the right portion of the domain ($x>15$ m).
The values of the conservative variables for the left and right states (in double precision) are respectively 
\begin{equation}
	\W_L^0 = 	\begin{sistema}
							h_L =  2.0\, \text{m}\\
							q_L =  0.5\, \text{m$^3$/(s$\,$m)}\\
							z_L =  0.0\, \text{m}\\
				\end{sistema}\,\quad\text{and}\quad
	\W_R^0 =	\begin{sistema}
							h_R =  2.0\, \text{m}\\
							q_R =  4.40526631244211\, \text{m$^3$/(s$\,$m)}\\ 
							z_R = -0.14000491636663\, \text{m}\\							
						\end{sistema}\,.
\label{eq:murillo}
\end{equation}
The test is performed with $\xi=1$ and $A_g =0.01$ s$^2$/m for the Grass formula \eqref{eq:Grass}; $\epsilon=1\times10^{-4}$ is the selected value for the small parameter in the PRICE-C method.
The final time of the simulation is $t_\mathtt{end}=1.5$ s.

For the convergence analysis, two different sets of computational grids are considered: applying the original DOT method the channel is discretized with $\mathtt{nc}=$ 100, 250, 500, 750 and 1000 cells; using the more efficient PRICE-C method and A-DOT formulation, the domain is divided into 500, 750, 1000, 1500 and 3000 cells.

Figure \ref{fig:murillo} shows the results of the efficiency test.
Panels \emph{a, b} and \emph{c} show the comparison between analytical and numerical solutions at the end of the simulation in terms of conservative variables, with the coarsest grid of each set chosen to compare solutions with similar computational costs ($\mathtt{nc}=$ 100 for the DOT scheme and $\mathtt{nc}=$ 500 for the PRICE-C and A-DOT schemes, which correspond to a CPU time of about 0.1 second for all three solutions).
These panels show that all the three numerical schemes well fit the analytical solution, even though the PRICE-C and A-DOT solutions are rather closer to the analytical one.

In terms of numerical efficiency, according to panels \emph{d, e} and \emph{f} of Fig.~\ref{fig:murillo}, all three methods converge to the exact solution with the same rate.
Indeed, for each method, the computed errors lie approximately on straight lines parallel to each other.
This behaviour is shown by all three methods and for all the conservative variables, verifying an overall convergence with the expected first order of accuracy.

In support of the better performance of PRICE-C and A-DOT, the right panels of Fig.~\ref{fig:murillo} indicate that the original DOT is always the least efficient method, while PRICE-C falls between DOT and A-DOT, the last one being the most efficient.
Considering only the CPU time, for a given computational grid, PRICE-C is the fastest method.
Otherwise, in terms of numerical efficiency, it behaves differently depending on the considered variable: it reproduces the hydrodynamics very well (with a performance close to A-DOT), whereas the solution for the bed elevation is close to the DOT solution.
Thus, for this test case, the A-DOT model is the best choice: given a target error (i.e.\ a fixed value for $\Eg$), A-DOT is constantly about ten times faster than the original DOT; when the CPU time is fixed, it is always the most accurate method.

\subsection{Test 2: smooth analytical solution}
\label{sub:berthon}
\begin{figure}[tp]%
	\centering
	\includegraphics[width=1\columnwidth]{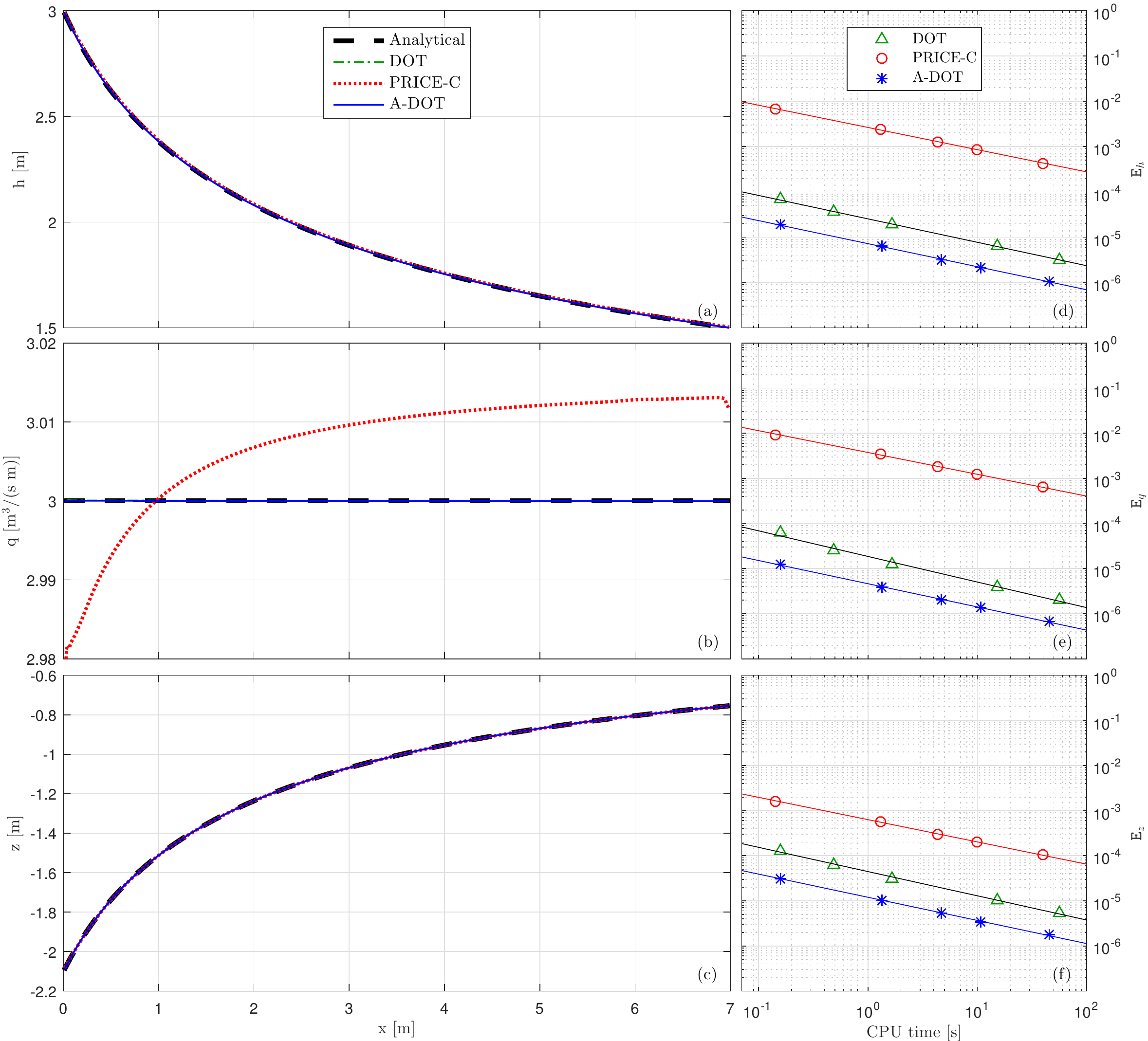}%
	\caption{Results of Test 2. Panels \emph{a, b} and \emph{c}: comparison between analytical, Eq.~\eqref{eq:berthon}, and numerical solutions (for $h$, $q$ and $z$, respectively) at the end of the simulation (with grid size $\mathtt{nc}=$ 100 for the DOT scheme; $\mathtt{nc}=$ 300 for the PRICE-C and A-DOT schemes).  Panels \emph{d, e} and \emph{f}: CPU time versus  $\Eg$, Eq.~\eqref{eq:Eg}, with $\varphi = h$, $q$ and $z$, respectively, by using: $\nc=[25, 50, 100, 300, 600]$ for DOT; $\nc=[100, 300, 600, 900, 1800]$ for PRICE-C and A-DOT.}%
	\label{fig:berthon}%
\end{figure}
The smooth analytical solution refers to a steady-state condition for a subcritical water flow coupled with a linear-in-time bed erosion, as proposed by \citet{Berthon2012}.

For a given uniform discharge $q_0$ and for a sediment transport assumed to be a power function of the velocity \eqref{eq:Grass}, the analytical solution of \eqref{eq:quasiLinNC} is given by
\begin{equation}
	\left\{
	\begin{aligned}
		&u(x) = \left[\frac{\alpha\, x+\beta}{A_g}\right]^{1/3} \Rightarrow\;\; h(x) = q_0/u(x)\\
		&z(x,0)= -\frac{u(x)^3+2\,g\,q_0}{2\,g\,u(x)} + \mathtt{C}\\
		&z(x,t)=-\alpha\, t+z(x,0)
	\end{aligned}
	\right.
	\label{eq:berthon}
\end{equation}
in which the constants $\alpha$, $\beta$ and $\mathtt{C}$ are $\alpha=\beta=A_g=0.005$ s$^2$/m and $\mathtt{C}=1$ m. The proper initial conditions are:
\begin{equation}
	q(x,0)=q_0,\qquad
	h(x,0)=q_0 \left[ \frac{\alpha\,x+\beta}{A_g} \right]^{-1/3},\qquad
	z(x,0)=\mathtt{C} - \frac{ \frac{\alpha\,x+\beta}{A_g}  + 2\,g\,q_0}
	{2\,g\,\left( \frac{\alpha\,x+\beta}{A_g} \right)^{1/3}},
\end{equation}
while the boundary conditions, for the domain defined by $0\leq x \leq L$, are:
\begin{equation}
	q(0,t)=q_0,\qquad
	h(L,t)=q_0 \left[ \frac{\alpha\,L+\beta}{A_g} \right]^{-1/3},\qquad
	z(0,t)=\mathtt{C} - \frac{ \frac{\beta}{A_g}  + 2\,g\,q_0}
	{2\,g\,\left( \frac{\beta}{A_g} \right)^{1/3}} - \alpha\,t.
\end{equation}

In this numerical test, the analytical solution \eqref{eq:berthon} is applied to reproduce a 10-second evolution in a 7 m-long straight channel.
As in the previous test, the domain is discretized with two different sets of grids: $\mathtt{nc}=$ 25, 50, 100, 300 and 600 cells for the DOT model; $\mathtt{nc}=$ 100, 300, 600, 900 and 1800 cells for the A-DOT and PRICE-C models.
We used $\epsilon=1\times10^{-4}$ for PRICE-C, calibrating the model in order to minimize the normalised root square error in terms of bed elevation $\mathtt{E}_z$.

Panels \emph{a, b} and \emph{c} in Figure \ref{fig:berthon} show the comparison between analytical and numerical solutions at the end of the simulation.
This comparison is presented using $\mathtt{nc}=$ 100 for DOT and $\mathtt{nc}=$ 300 for PRICE-C and A-DOT, corresponding to a CPU time of about 1.3 seconds for all three simulations.

Figure \ref{fig:berthon}\emph{b} highlights a different behavior between the two versions of the DOT scheme and PRICE-C.
This scheme is not able to reproduce the constant water discharge as the first two methods.
Thus the whole PRICE-C solution is affected by a larger mean root square error, which also decreases the efficiency of the method in this particular test.

The difference between the PRICE-C and DOT schemes is quantified by the efficiency comparison (in the right panels of Fig.~\ref{fig:berthon}):
with constant CPU time, A-DOT is almost 100 times more accurate in terms of bed elevation and 1000 times more accurate in terms of water depth and liquid discharge than PRICE-C.
In spite of this marked difference, the rate of convergence of all the methods is still the same (as shown by the same slope of the fitting lines).
Moreover, the mean errors measured for PRICE-C are still of the same order of magnitude as the same errors in the previous test, while DOT and A-DOT appear very much more accurate.

These results are due to the quasi-steadiness of the flow.
In this condition, the bed evolution is mainly related to the intermediate wave associated with the smallest eigenvalue of the dSVE equations.
Therefore, for the considerations made in section \ref{sub:DOT}, it is not surprising that a complete Riemann solver like DOT, performs better than an incomplete Riemann solver like PRICE-C.

Focusing only on the comparison between the DOT and A-DOT methods, Figures \ref{fig:berthon}\emph{d, e} and \emph{f} confirm that the analytical computation of $|\A|$ produces a ten-time increase of the CPU performance, which remains the same for all three conservative variables.
Thus, also in this test, the performance of the new implementation with respect to the original DOT increases by the same factor as in Test 1.

\subsection{Test 3: subcritical and supercritical flow over a movable hump}
\label{sub:cordier}
\begin{figure}[tp]%
	\centering
	\includegraphics[width=1\columnwidth]{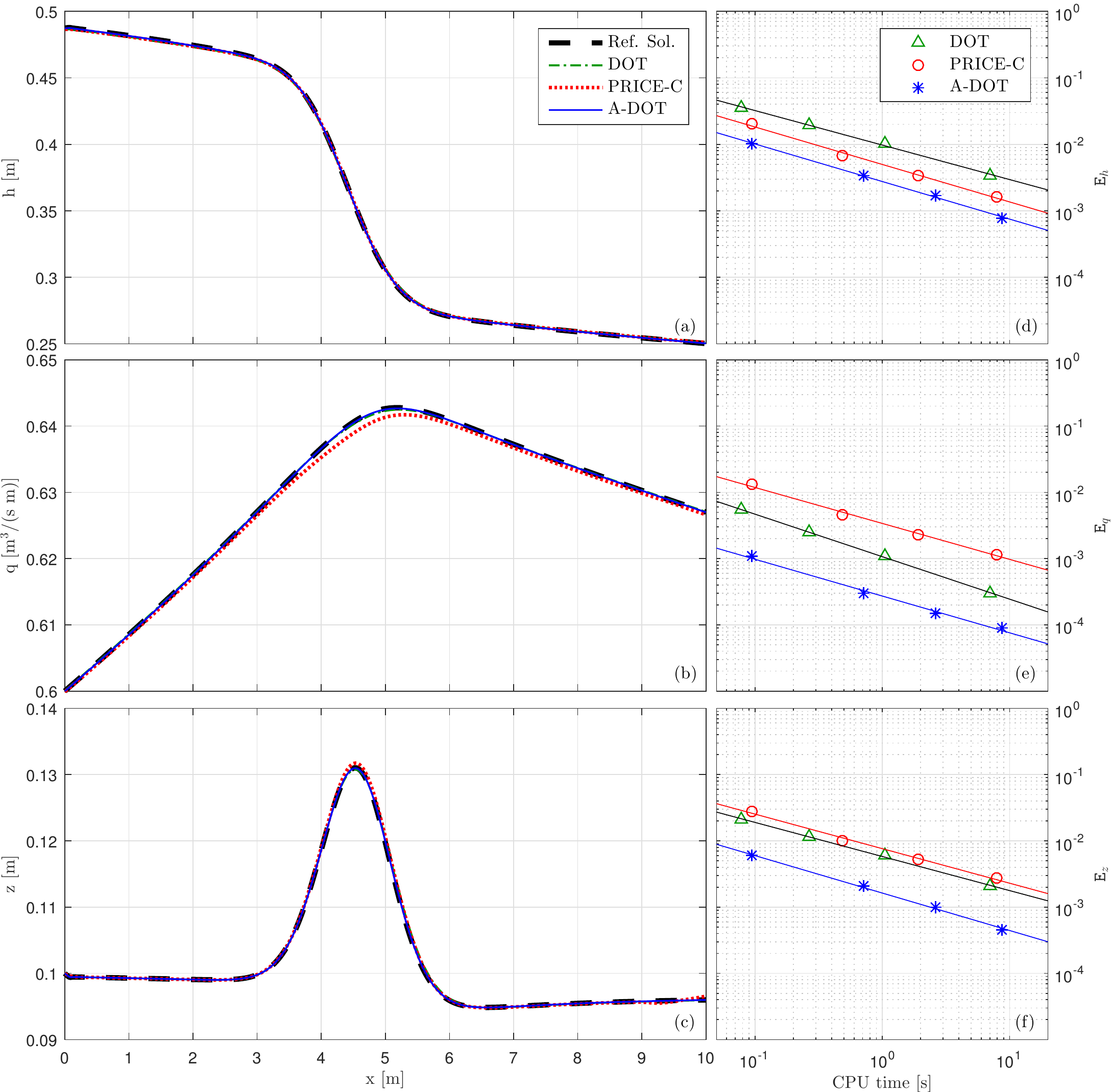}%
	\caption{Results of Test 3. Panels \emph{a, b} and \emph{c}: comparison between reference and numerical solutions (for $h$, $q$ and $z$, respectively) at the end of the simulation (with grid size $\mathtt{nc}=$ 300 for the DOT scheme; $\mathtt{nc}=$ 1200 for the PRICE-C and A-DOT schemes).  Panels \emph{d, e} and \emph{f}: CPU time versus  $\Eg$, Eq.~\eqref{eq:Eg}, with $\varphi = h$ $q$, and $z$, respectively, by using: $\nc=[25, 50, 100, 300]$ for DOT; $\nc=[100, 300, 600, 1200]$ for PRICE-C and A-DOT.}
	\label{fig:transC}%
\end{figure}
The morphodynamic evolution of a bed hump is considered for the last efficiency test.
A water stream with subcritical and supercritical regions flows over the hump, in a 10 m-long channel.
This numerical application is inspired by the test proposed by \citet{Cordier2011} to prove that the dSVE equations must be solved with fully coupled models to avoid non-physical instabilities of the solution.

A preliminary steady-state solution over a fixed bed ($A_g = 0.0$) is computed starting from the following initial condition:
\begin{equation}
	\left\{
	\begin{aligned}
		&q(x,0) = 0.6 \;\text{m}^3/\text{s m},\\
		&z(x,0)= 0.1 + 0.1\,e^{-\left(x-5\right)^2} \;\text{m}\\
		&h(x,0) + z(x,0) = 0.4 \;\text{m}
	\end{aligned}
	\right.\,,
	\label{eq:cordier}
\end{equation}
when a constant liquid discharge at the upstream boundary and transmissive conditions at the downstream boundary are imposed, the fixed bed simulation is carried out until the (frictionless) steady solution is reached.
Then the bed is allowed to evolve by using $A_g$ = 0.0005 s$^2$/m in the Grass bed-load formula \eqref{eq:Grass}. 

The efficiency of the methods is evaluated in a morphodynamic evolution of 10 seconds, taking as reference the numerical solution provided by the PRICE-C method with a very refined computational grid ($\nc = 7200$).
The tested grids are: $\nc=$ 25, 50, 100, 300 cells for DOT; $\nc=$ 100, 300, 600 and 1200 cells for A-DOT and PRICE-C.
For PRICE-C, $\epsilon=1\times10^{-2}$ is used, which is higher than in the previous tests in accordance with the faster bed adaptation due to the near-critical water flow conditions.

In Figure \ref{fig:transC}, panels \emph{a, b} and \emph{c} show the comparison between reference and test solutions at the end of the simulation, corresponding to a CPU time of about 10 seconds for all three solutions, with $\nc=$ 300 for the DOT scheme and $\nc=$ 1200 for the PRICE-C and A-DOT schemes.
The three methods well approximate the reference solution.
Only PRICE-C presents a small difference in terms of liquid discharge where the water flow crosses the bed hump under critical conditions.
Nevertheless, this can be considered a good performance for all the methods, taking into account the critical hydrodynamic conditions.

The efficiency study presented in panels \emph{d, e} and \emph{f} of Fig.~\ref{fig:transC} shows that, except for the liquid discharge evaluation, the numerical performance of PRICE-C is very good.
Notwithstanding this, the A-DOT method still provides the best numerical efficiency in this test.
Indeed, the new analytical implementation is very robust even when the water flow is characterized by critical conditions, maintaining the same performance observed in the previous two tests.
In other words, A-DOT can reproduce the same solution as the original DOT, but with the advantage of being ten times faster.
Moreover, the A-DOT solver is also more efficient than the PRICE-C solver in the computation of all three conservative variables.

\section{Comparison with observed data}
\label{sec:experimental}
The proposed A-DOT numerical model, once validated on theoretical benchmarks, is here tested in a practical context.
First, a case study involving a progressive channel aggradation due to sediment overfeeding \cite{Soni1980117}  allows to show the advantage of the A-DOT model with respect to the PRICE-C model in terms of numerical efficiency.
Then, the propagation of a sediment bore within a flume, experimentally investigated by \citet{Bellal2003} and already used as benchmark by \citet{Canestrelli2009}, is reproduced using different transport formulae, to show the independence of the present approach from the sediment transport closure equations.

\subsection{Aggradation due to overloading}
\label{subs:Soni}
The first test case consists of a bed aggradation in a laboratory flume due to sediment overfeeding at its upstream boundary \cite{Soni1980117}. In nature, this kind of aggradation process may be caused by a hill-side landslide into a river.
The laboratory setup consisted of a 0.5 m deep and 30 m long tilting flume, equipped with a recirculatory device. The width of the flume is $B=0.2$ m.
The sediment forming the bed was sand with a median diameter of 0.32 mm and geometric standard deviation of 1.3.

Experiments started imposing a constant water discharge $Q_0$ at the upstream section of the channel until equilibrium sediment discharge $Q_{S0}$, bed slope $s_0$ and uniform water depth $h_0$ were established.
This steady condition was perturbed by increasing the upstream sediment feeding from $Q_{S0}$ to $Q_{S0}+\Delta Q_S$.
As a result, the riverbed steepened leading to a progressive channel bed aggradation.

\citet{Soni1980117} performed several experimental runs changing the water discharge, the equilibrium bed slope and the sediment overfeeding amount $\Delta Q_S$.
Here, the most intense case of aggradation is considered to compare numerical results of the A-DOT and PRICE-C schemes with experimental data. In such case $\Delta Q_S = 4\,Q_{S0}$ and $Q_0 = 0.004\;\text{m}^3/\text{s}$. 
The numerical simulations are performed considering the unit width of the flume and therefore the initial conditions for the numerical runs are:
\begin{equation}
	\left\{
	\begin{aligned}
		&h(x,0) = h_0 = 0.05\; \text{m}\\
		&q(x,0) = q_0 = Q_0/B= 0.02 \;\text{m}^3/\text{s m}\\
		&z(x,0) = 1.2 - x\,s_0 \;\; \text{m}\quad \text{with}\quad s_0 = 3.56 \text{\textperthousand}
	\end{aligned}
	\right.\,.
	\label{eq:Soni0}
\end{equation}

Constant specific liquid discharge $q_0$ and specific solid discharge $q_s = (Q_{S0}+\Delta Q_S)/B$ are set at the upstream boundary, while constant water depth equal to $h_0$ is imposed at the downstream boundary.

The solid discharge during the simulation is evaluated by a power function of velocity, similar to the Grass formula \eqref{eq:Grass}, using the original parameters experimentally derived by \citet{Soni1980117}:
\begin{equation}
	q_s = \alpha \, u^{\beta} \quad\text{with}\quad\alpha=1.45\times10^{-3}\,,\quad\beta=5\,.
	\label{eq:Soni_qs}
\end{equation}
Bed porosity is assumed to be 0.4, the Strickler coefficient to be $K_s = 49.4$ m$^{1/3}$ s$^{-1}$; the computational domain is discretized with 100 cells.
Finally, the simulation time is set to $\mathtt{T_{end}}= 2400$ s.

By imposing the mass conservation of the solid phase, at the end of the simulation the aggradation of the bed at the inflow boundary results $\Delta z(0,\mathtt{T_{end}}) = 0.0675$ m.

According to \citet{Soni1980117}, the final experimental bed aggradation profile, $\Delta z(x,\mathtt{T_{end}})$, is well approximated by the following empirical relationship:
\begin{equation}
	\frac{\Delta z(x,\mathtt{T_{end}})}{\Delta z(0,\mathtt{T_{end}})} = 1-\text{erf}\left(\frac{x}{2\,\sqrt{K\,t}}\right)\,,
\label{eq:SoniZ}
\end{equation}
where $K$ is the aggradation coefficient which reads: 
\begin{equation}
	K= \frac{1}{\xi}\,\frac{\Delta Q_S}{\Delta z(0,\mathtt{T_{end}})} \;1.143\times10^{-3}\,.
\label{eq:Soni_K}
\end{equation}

\begin{figure}[tb]%
\centering
\includegraphics[width=.8\columnwidth]{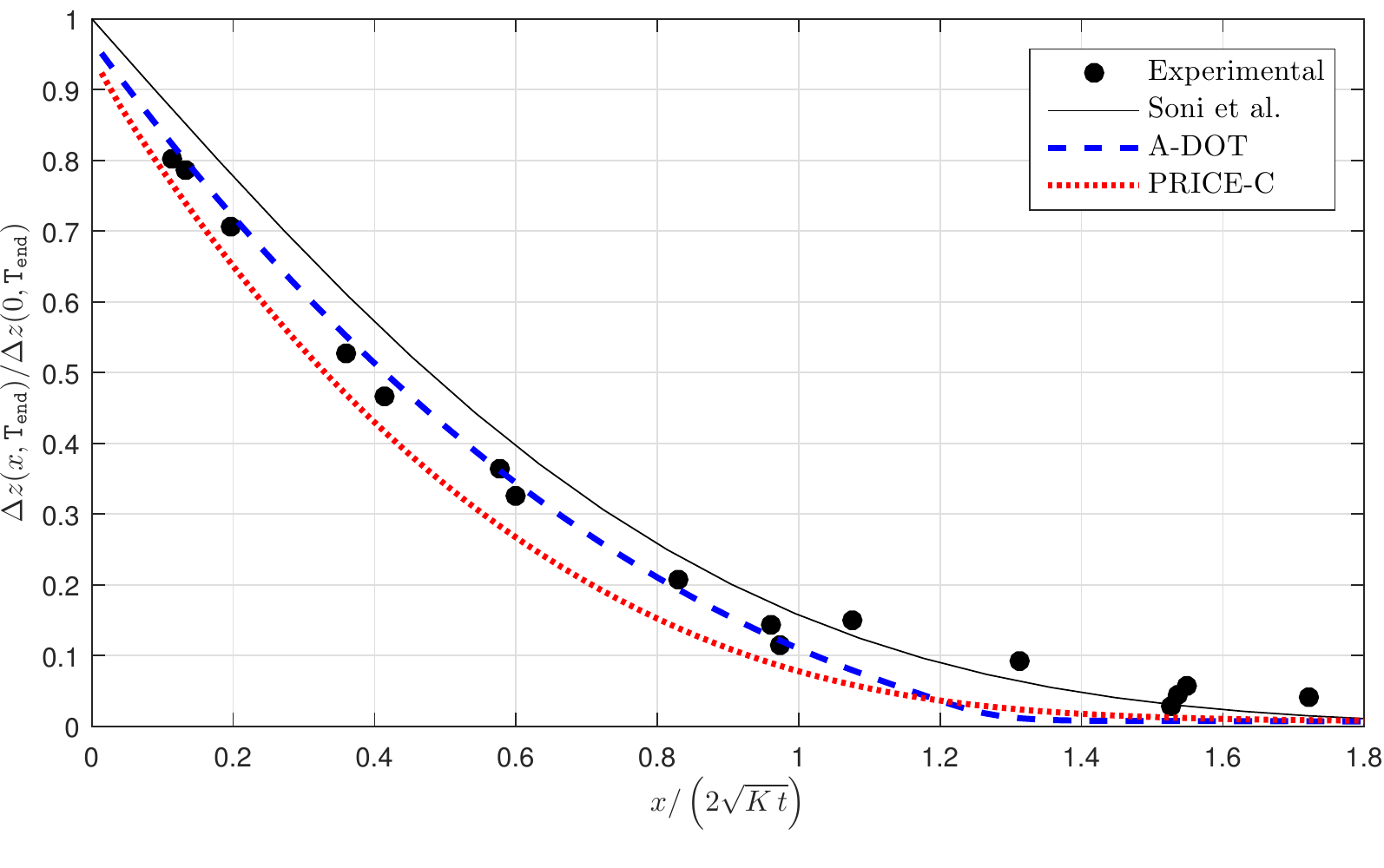}%
\caption{Non-dimensional bed aggradation due to sediment overfeeding of $\Delta Q_S = 4\,Q_{S0}$ at the inflow boundary, evaluated at the final time $\mathtt{T_{end}}=2400$ s: comparison between experimental data, analytical solution (Eq.\ \eqref{eq:SoniZ} by \citet{Soni1980117}) and numerical results of PRICE-C and A-DOT with $\nc=100$. Numerical simulations start from equilibrium conditions, Eqs.\ \eqref{eq:Soni0} and \eqref{eq:Soni_qs}; constant liquid discharge is imposed at the upstream boundary and uniform flow depth at the downstream section.}%
\label{fig_SoniBed}%
\end{figure}
In Fig.\ \ref{fig_SoniBed}, the A-DOT and PRICE-C numerical results are compared with the experimental data and the empirical relationship \eqref{eq:SoniZ}.
The numerical results obtained with the original DOT are not presented because they coincide with those obtained with the A-DOT.
According to Fig.\ \ref{fig_SoniBed}, the A-DOT numerical solution well fit experimental data, also better than the analytical function \eqref{eq:SoniZ}. 
To reproduce 2400 s of bed aggradation the A-DOT model needs 1.50 s, instead of 14.97 s of CPU time for the original DOT and 1.48 s for the PRICE-C model.
For the same discretization of the computational domain and very similar CPU time, the A-DOT solution predicts the bed elevation better than the PRICE-C (empirically calibrated with $\epsilon=0.05$).

\begin{figure}[tb]%
\includegraphics[width=\columnwidth]{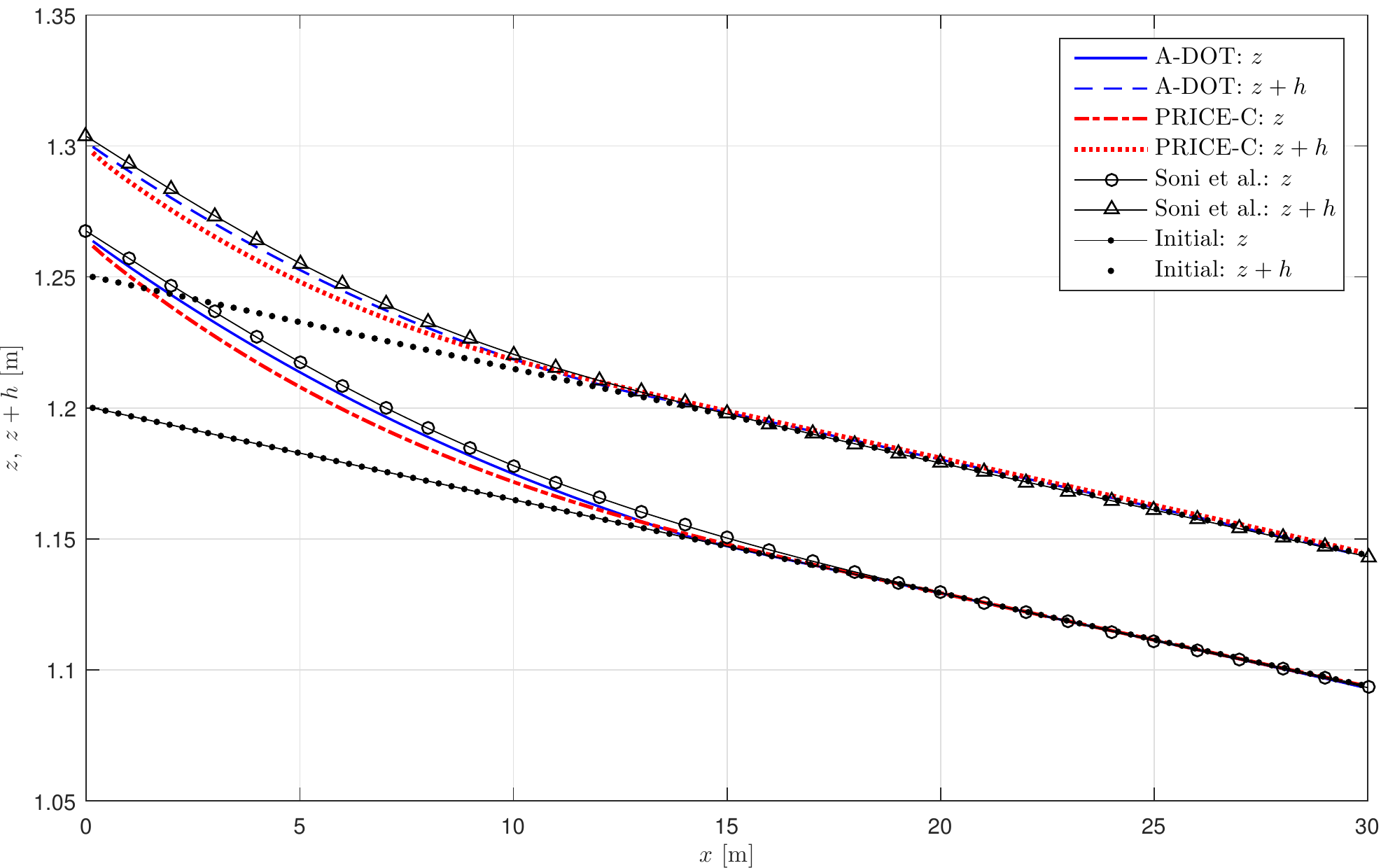}%
\caption{Bed and free surface profiles at the final time $\mathtt{T_{end}}=40$ minutes: comparison between reconstructed analytical solution (Eq.\ \eqref{eq:SoniZ} by \citet{Soni1980117}) and numerical results of PRICE-C and A-DOT with $\nc=100$. Numerical simulations start from equilibrium conditions of Eqs.\ \eqref{eq:Soni0} and \eqref{eq:Soni_qs}, while constant liquid discharge and sediment overfeeding of $\Delta Q_S = 4\,Q_{S0}$ are imposed at the upstream boundary and uniform flow water depth is imposed at the closure section.}%
\label{fig_SoniAll}%
\end{figure}
Given the analytical approximation for the bed aggradation \eqref{eq:SoniZ}, imposing constant liquid discharge and the uniform depth at the downstream boundary, the free surface profile at $t=\mathtt{T_{end}}$ is easily reconstructed.
This reconstruction is possible assuming that the aggradation process is slow enough that the bed evolution is weakly coupled to the hydrodynamic, such that the hydrodynamic profile can be approximated with the steady water flow over an ``instantaneous fixed bed''.
In Fig.\ \ref{fig_SoniAll} numerical results are compared with this reference solution.
Both numerical solutions well reproduce the aggradation phenomenon, but the A-DOT is closer to the reference solution than the PRICE-C, for the same spatial resolution $\nc=100$. Both methods produce very similar CPU time usage.

\subsection{Propagation of a sediment bore}
\label{subs:Toro}
The propagation of a sediment bore is experimentally investigated by \citet{Bellal2003}.
Here, their experiments are considered to test the effect of different transport formulae on the A-DOT numerical results, similarly to what has been done in \cite{Canestrelli2009} for the PRICE-C scheme.

The experimental set up consists in a steep-sloped ($s_0 = 3.02\%$), 6.9 m long and 0.50 m wide, rectangular flume. The bottom is covered with a uniform coarse sand with a mean size of 1.65 mm and porosity of 0.42.
Initially, the flume is fed with constant water discharge $Q=12.0$ l/s and sediment discharge $q_s=0.196$ l/s until the bed profile has reached a quasi-equilibrium conditions.
At the time t = 0, at the downstream end of the flume, the rapid raise of a submerged weir perturbs the equilibrium situation by imposing a subcritical condition at the closure section.
The induced water level at the downstream end is $H = 20.93$ cm, while both water and sediment discharges at the upstream section are kept constant for the entire duration of the experiment.
The selected Strickler coefficient is $K_s=62$ m$^{1/3}$ s$^{-1}$.

The described hydraulic configuration gives rise to a moving hydraulic jump and a sediment bore which represents a demanding test case for numerical schemes.
For example, they may fail in predicting the propagation celerity (i.e. the position in time) of such bore caused by the presence of transcritical flow.

Numerical simulations are performed by using the sediment transport formulae \eqref{eq:MPM}, \eqref{eq:vanRijn} and the power-law \eqref{eq:GrassThr}.
In particular, Eq.\ \eqref{eq:GrassThr} is used with the parameters defined in \cite{Canestrelli2009}, i.e. $A_g = 0.0024$ s$^2$/m and $u_{cr} = 0.3$ m/s.
For all numerical simulations the same computational grid, with $\nc=200$, is used.

\begin{figure}[t]%
\centering
\includegraphics[width=0.8\columnwidth]{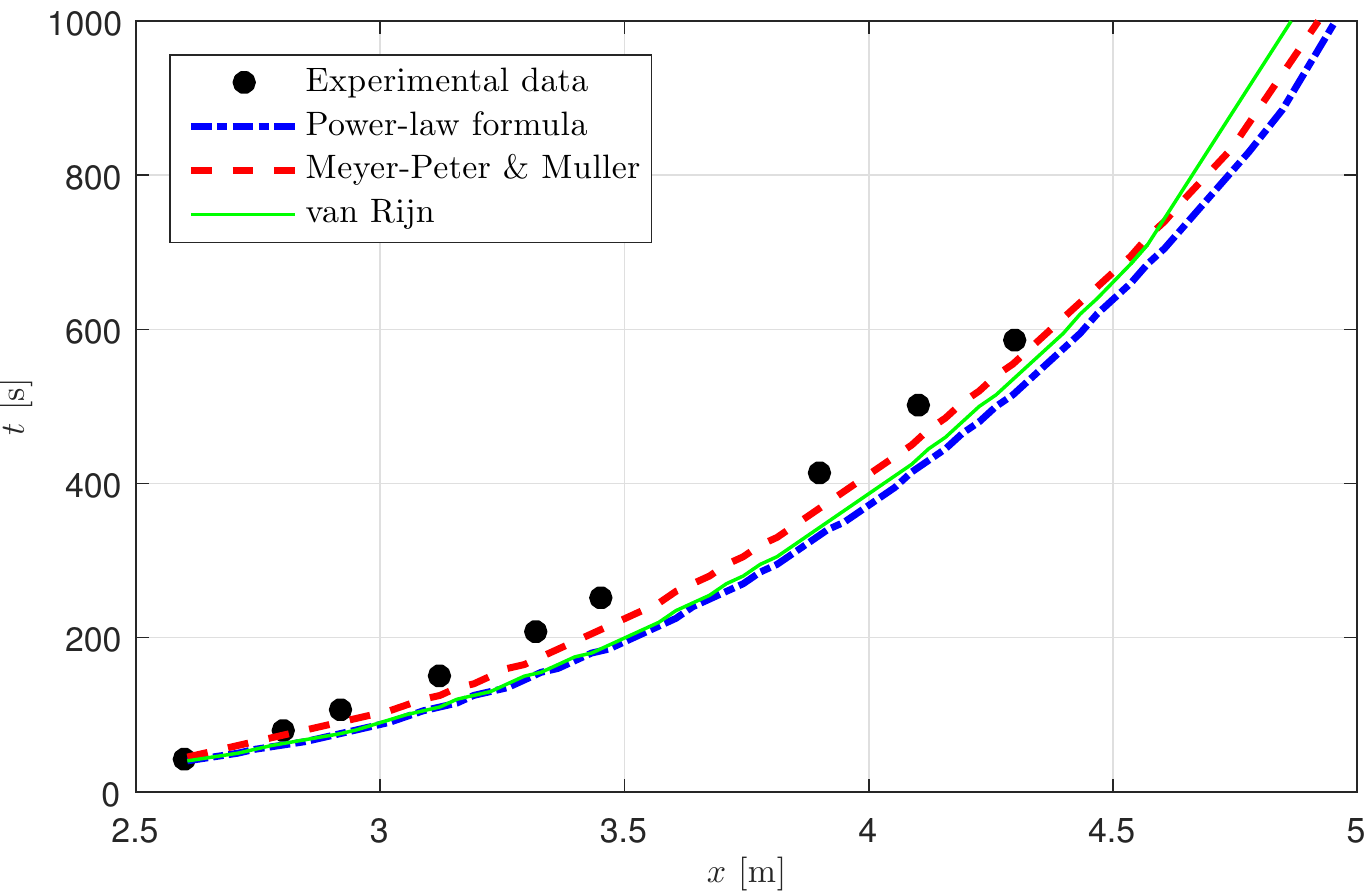}%
\caption{Front positions: comparison between experimental data and A-DOT scheme using three different formulae for the solid discharge.}%
\label{fig:DatiToro}%
\end{figure}
According to \citet{Canestrelli2009}, the shock position at time $t$ is conventionally assumed coincident with the barycentre coordinate of the first cell (starting from the downstream boundary) that satisfies:
\begin{equation}
	z_i^{n+1} - z_i^n >\tau
\label{eq:ToroFrontX}
\end{equation}
where $\tau$ is a given tolerance fixed to 0.02 m in all numerical tests.
In Fig.\ \ref{fig:DatiToro} the position of the sediment bore is plotted as a function of time for all the considered sediment discharge formulae and it is compared with the experimental data.
The numerical results are closed to the experimental data and are also consistent with those obtained by \citet{Canestrelli2009}.
According to Fig.\ \ref{fig:DatiToro}, the bore celerity depends on the sediment transport formula.
These results demonstrate that the proposed numerical model can be used with different transport formulae, leaving unchanged the general structure of the method.

\section{Conclusions}
\label{sec:conclusion}
Within the framework of the de Saint Venant-Exner model, an analytical formulation of the flux matrix eigenstructure is used to compute the jump functions at the cell edges to improve the numerical performance of the Dumbser-Osher-Toro path-conservative scheme.
A study of the numerical efficiency and a strong validation of the A-DOT scheme is presented, using as references original DOT and PRICE-C schemes and two different experimental data sets.

With the computational grid resolution fixed, the A-DOT scheme computes the solution ten times faster than DOT, keeping unchanged the accuracy of the solution itself.
Moreover, this efficiency boost does not depend on a specific test or particular morphodynamic conditions.
Similarly, with a target accuracy fixed, the A-DOT scheme also presents a better performance than the efficient PRICE-C scheme.

The computational time becomes a crucial aspect when long-term morphological changes of river, estuarine and coastal environments must be studied in engineering/planning applications.

The high performance of the new A-DOT scheme is also achieved by simplifying the implementation of the method.
While the original DOT scheme requires the use of external numerical libraries to compute the eigenstructure of the system matrix $\A$, the A-DOT jump function is merely computed as the product of three known matrices. 
Thus, the A-DOT scheme is both easy to implement and very portable within different coding environments and languages.
Furthermore, in the framework of the de Saint Venant-Exner model, the A-DOT method is fully compatible with any kind of sediment transport formula, providing that the system of PDE remains strictly hyperbolic.

Except for the proper closure for the solid discharge and the friction factor (as typical of each movable bed problem), the method does not require any parameter calibration. Therefore, any unphysical tuning of diffusion is avoided obtaining a complete reliability and reproducibility.

The idea to increase the computational efficiency of the DOT scheme by analytically computing the eigensystem is presented only for the de Saint Venant-Exner mathematical model.
Nonetheless, it can also be applied to enhance the DOT performance with other mathematical models.
Moreover, for the sake of simplicity, we applied the A-DOT method to a first-order path-conservative scheme using a simple segment path, although there are no restrictions either on the order of accuracy of the method or on the path definition.

\section*{Funding} 
The first author (F. Carraro) is funded by the University of Ferrara through the 5\textperthousand~ fees donation within the ``Young Researcher Project''.

The second author (A. Valiani) is funded by the University of Ferrara within the Founding Program FAR 2016.

The third author (V. Caleffi) is funded by the University of Ferrara within the Founding Program FIR 2016, project title ``Energy-preserving numerical models for the Shallow Water Equations''.

\section*{References}
\bibliographystyle{elsarticle-harv}
\bibliography{ADOTvsPRICE}


%

\end{document}